\documentclass[%
 aip,
 floatfix,
 amsmath,amssymb,
 reprint,%
]{revtex4-1}

\usepackage{rotating}
\usepackage{graphicx}
\usepackage{dcolumn}
\usepackage{bm}
\usepackage{soul,xcolor}
\usepackage{siunitx}
\usepackage{textcomp}
\setstcolor{red}
\RequirePackage{color}
\usepackage[utf8]{inputenc}
\usepackage[T1]{fontenc}
\usepackage{mathptmx}
\usepackage[normalem]{ulem}

\begin{document}

\preprint{AIP/123-QED}

\title[Resonant transport in Kekulé -distorted graphene nanoribbons ]{Resonant transport in Kekulé-distorted graphene nanoribbons}

\author{Elias Andrade}

\affiliation{%
 Facultad de Ciencias, Universidad Aut\'{o}noma de Baja California,
Apdo. Postal 1880, 22800 Ensenada, Baja California, M\'{e}xico. 
}%

\author{Ramon Carrillo-Bastos}
\email{ramoncarrillo@uabc.edu.mx}
\affiliation{%
 Facultad de Ciencias, Universidad Aut\'{o}noma de Baja California,
Apdo. Postal 1880, 22800 Ensenada, Baja California, M\'{e}xico. 
}%
\author{Pierre A. Pantale\'{o}n}
\affiliation{School of Physics and Astronomy, University of Manchester, Manchester M13 9PL, United Kingdom}
\affiliation{IMDEA Nanociencia, C. Faraday 9, 28015 Madrid, Spain }

\author{Francisco Mireles}
\affiliation{Departamento de F\'isica, Centro de Nanociencias y Nanotecnolog\'ia, Universidad Nacional Aut\'onoma de M\'exico, Apdo. Postal 14, 22800 Ensenada, Baja California, M\'exico}
\date{\today}

\begin{abstract}
The formation of a superlattice in graphene can serve as a way to modify its electronic bandstructure and thus to engineer its electronic transport properties. Recent experiments have discovered a Kekul\'{e} bond ordering in graphene deposited on top of a Copper substrate, leading to  the breaking of the valley degeneracy while preserving the highly desirable feature of linearity and gapless character of its band dispersion. In this paper we study the effects of a Kekul\'{e} distortion in zigzag graphene nanoribbons in both, the subband spectrum and on its electronic transport properties. We extend our study to investigate also  the electronic conductance in graphene nanoribbons composed of sequentially ordered $\nu=\pm1$ Kek-Y superlattice. We find interesting resonances in the conductance response emerging in the otherwise energy gap regions, which scales with the number of Kek-Y interfaces minus one. Such features resembles the physics of resonant tunneling behavior observed in semiconductors heterostructures. Our findings provide a possible way to measure the strenght of Kekul\'{e} parameter in graphene nanoribbons.
\end{abstract}

\maketitle

\section{\label{sec:intro}Introduction}
The incorporation of graphene nanoribbons is the natural route to bring graphene special properties to modern day electronics, as they are expected to be used \textit{e.g.} in graphene-based solid and flexible devices, as well as in its wiring connections \cite{bischoff2015localized}. The electronic transport properties of these narrow graphene stripes have been widely studied\cite{brey2006electronic,NakadaGNR, FujitaGNR,AkhmerovGNR, wakabayashi2010electronic,bischoff2015localized}.  
The main caveat of such electronic devices is the lack of an intrinsic band gap in graphene, which results for instance in poor current on/off ratios in field effect transistors 
\cite{bischoff2015localized,roldan2015strain}. There are a number of proposals to circumvent this issue: confinement in quasi-one-dimensional systems such as graphene nanoribbons (GNRs)\cite{wakabayashi2010electronic}, inducing strain effects\cite{pereira-GNR,vozmediano2010gauge, AMORIM20161, NaumisReview, Bahamon}, doping graphene with impurities\cite{filho2007,Usachov2016}, and exploiting spin-orbit coupling mechanisms\cite{KaneyMele,SpinorbitGAP,Magneto-SO,Proximity-SO}, just to mention a few. 
 
Yet  another appealing proposal  for  opening  a  band  gap  in graphene  is  through the inclusion of a Kekul\'e-type
distortion\cite{chamon2000solitons,cheianov2009hidden,manoharan2012,gamayun2018valley}.  It  consists in modifying the bond strength of electrons in the $\pi$-band in an alternating pattern (with two distinct carbon-carbon bond lengths), in such a way that the translational symmetry is reduced respect to pristine graphene, resulting in a primitive cell of six carbon atoms instead. As a consequence, the new  Brillouin zone can be seen as a folding down the original  Brillouin zone to the $\Gamma$-point having now the two valleys degenerated at its center\cite{Ren2018}. The pattern in which the carbon-carbon bond strength is altered as in a benzene ring is known as Kek-O texture. This texture has the property of opening a band gap, and can be seen as a two dimensional extension of the Peierls metal-insulator transition in 1D lattices \cite{chamon2000solitons,lee2011band}. On the other hand, the texture in which the three modified neighboring bonds form a  Y-shape pattern (\textit{i.e.} surrounding one of the carbon atoms of the new hexagonal unit cell) is called Kek-Y distortion. In contrast with the Kek-O, the Kek-Y shape does not open a gap, but instead, it leads to the locking of valley degree of freedom to the momentum\cite{gamayun2018valley}.

 Since the seminal theoretical work by C. Chamon\cite{chamon2000solitons}, several authors have studied the physical consequences of a Kek-O distortion in graphene. For instance it has been pointed out that a Kek-O distortion  allows the realization of charge fractionalization with time reversal symmetry in 2D graphene-like systems\cite{Chamon2007,chamon2008electron},  leading to the formation of states with topological properties\cite{kariyado2017topological,liu2017topological}. There is no experimental evidence yet of the Kek-O phase in graphene\cite{cheianov2009hidden,lin2017competing}, although it can be achieved in  analogues systems like  molecular graphene\cite{manoharan2012} and vibrational systems \cite{Pseudospins2017}. As for  the  Kek-Y phase, it became a subject of scrutiny after its recent experimental realization by C. Gutierrez, \textit{et al}.\cite{gutierrez2016imaging}, whom showed amazing STM images of standing-wave in the local density of states of graphene on Cu(111) surface forming a Kek-Y pattern. Soon after, Gamayun \textit{et al}. \cite{gamayun2018valley} studied theoretically the low energy band dispersion laws of the Kek-Y distorted graphene finding a gapless spectrum with valley-momentum locking for the charge carriers. Other  recent theoretical studies of the Kek-Y phase in graphene describe its effects on Klein tunneling\cite{Juan-prb}, its interplay with mechanical strain\cite{Arraga2018,EliasPaper} and the formation of multiflavor Dirac Fermions in Kekul\'e-textured graphene bilayers \cite{DavidRuiz}. 
 
 In the experiments by Gutierrez \textit{et al.}\cite{gutierrez2016imaging} of graphene/Cu,  the observed formation of Kekul\'e ordering was inferred  from the interactions between vacancies in the Cu substrate and the carbon atoms of graphene. The authors report to observe large areas with the same orientation of Kek-Y distortion, however they also noticed that regions where two or more orientations of Kek-Y distortion were present as well. Until now, a theoretical study of the effects in such interfaces on the electronic transport has not been explored, nor the effect of the Kekul\'e-Y distortion in graphene nanoribbons on its subband spectrum and on its transport properties. 
 
  In this work we study the effects of Kekul\'e textures of both Y and O types on the subband structure of zigzag GNRs. Analytical results for the band structure show 
  unusual band gaps originated by a combined effect of the Kekul\'e texture and the zigzag boundary. Numerical calculations for the electronic conductance through finite regions of Kek-Y interfaces reveal the appearance of resonant transport in the absence of local gating or position dependent external potential. We also introduce a simple way to extract a measure of the strength of the Kekulé parameter in graphene nanorribons.
  
  The paper is organized as follows:  In Sec. \ref{sec:Model and methods} we introduce the model for an infinite graphene zigzag nanorribon with an uniform Kekul\'e distortion. In Sec. \ref{sec:Results} we plot and discuss the results, including the conductance calculations. Finally, we draw our conclusions in Sec. \ref{sec:conclusions}.

\section{Model and methods} \label{sec:Model and methods}
We first consider an infinite graphene nanoribbon of width $W$ 
with zigzag edges and characterized by an uniform Kekul\'{e} distortion.  
Following Gamayun \textit{et al.}\cite{gamayun2018valley}, the nearest neighbors tight-binding Hamiltonian describing such system reads,
\begin{equation}
H=\sum_{l}\sum_{j=1}^{3}t_{l,j}a_{\mathbf{r}_l}^{\dagger}b_{\mathbf{r}_l+ \bm \delta_{j}}+H.c.,
\end{equation}
where $l$ runs over all the sites of the sublattice A, $j$ runs over the three corresponding first-neighbors sites of the sublattice B connected through the vectors $ \bm \delta_1=a_{cc}(\frac{\sqrt{3}}{2},-\frac{1}{2})$, $ \bm \delta_2=a_{cc}(-\frac{\sqrt{3}}{2},-\frac{1}{2})$, $ \bm \delta_3=a_{cc}(0,1)$, where $a_{cc}=$1.42 \AA~is the carbon-carbon distance. Here $\mathbf{r}_{l}=n_l\mathbf{a}_1+m_l\mathbf{a}_2$ is the position vector of the $l$-th site in terms of the lattice vectors $\mathbf{a}_1=a(\frac{-1}{2},\frac{\sqrt{3}}{2})$
and $\mathbf{a}_2=a(\frac{1}{2},\frac{\sqrt{3}}{2})$ where $a=\sqrt{3}a_{cc}$ is the lattice parameter. For pristine graphene
the hopping integral between nearest-neighbor sites $t_{l,j}=t_{0}$  is around 2.7 eV. The Kekul\'e distortion of the lattice modifies the bond lengths and it is introduced through a position dependent hopping integral given by\cite{gamayun2018valley},
\begin{equation}
    t_{l,j}=t_0(1+2 \Re [\Delta e^{i(p \bm K_+ + q \bm K_-)\cdot \bm \delta_j+i \bm G \cdot \bm r_l}] ),
    \label{Eq:tij}
\end{equation}
where $\Delta=e^{i2 \pi N/3 } \Delta_0$ is the Kekul\'e coupling with amplitude $\Delta_0$ and a phase fixed by an arbitrary integer number $N$; typically is assumed that $|\Delta|\lesssim$0.1. In the above equation, $ \bm K_{\pm}=\frac{2 \pi}{9a}\sqrt{3}(\pm 1,\sqrt{3})$ and $\bm G=~\frac{4 \pi}{9a}\sqrt{3}(1,0)$ are the reciprocal lattice and the Kekul\'e wavevectors, respectively. The integers $p$ and $q$ 
define the type of Kekul\'e texture through the number
\begin{equation}
   \nu = (1+q-p)\mod 3,
\end{equation}
with $\{p,q\}\in \mathbb{Z}_3$ such that  a Kek-O bond texture corresponds to $\nu=0$, whereas the Kek-Y texture corresponds to $\nu=\pm 1$.
\begin{figure}[!htbp]
\begin{center}
\includegraphics[scale=0.31]{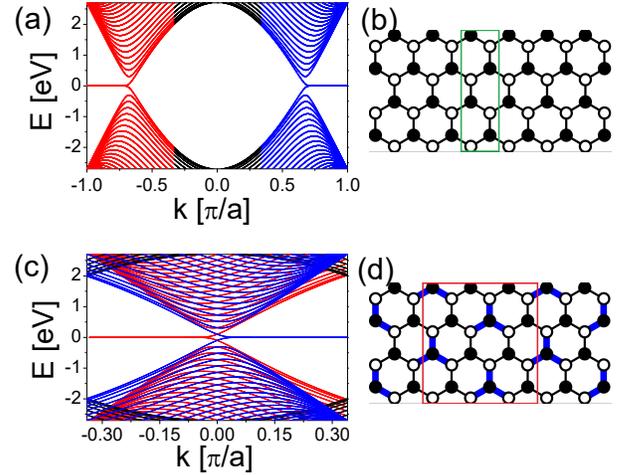}
\end{center}
\caption {(Color online) (a) Band structure for zigzag pristine graphene nanoribbons obtained with the green unit cell showed in (b). Band structure for zigzag graphene nanoribbons with Kek-Y distortion, obtained with the red unit cell showed in (d). 
\label{Fig1}}
\end{figure}

The aim is thus to calculate the subband spectrum of an infinite zigzag graphene nanoribbon with an uniform  Kek-Y and Kek-0 pattern. We then shall focus our study on the electronic properties (conductance) for Kek-Y superlattices. In order to tackle the subband spectrum problem and facilitate its numerical calculation it is necessary to define an appropriate unit cell\cite{datta2005}. Let $H_{00}$ be the matrix elements relating the sites within the same cell and $H_{01}$($H_{10}$) be the matrix elements relating the sites of that cell with the sites of its right(left) neighboring cell, see Fig. 1(b) and (d) for examples of unit cell chosen. It is not difficult to show that the subband structure for such periodic system can be obtained by solving the following secular equation,
\begin{equation}
\text{det}\bigg[ H_{00}+H_{01}e^{3ik_x b}+H_{10}e^{-3ik_xb}-E\mathbf{I} \bigg]=0,
\end{equation}
 for the energy $E(k_x)$ at each $k_x$ point within the first Brillouin zone. In the absence of the Kekul\'{e} distortion, a pristine zigzag GNR can be described with a unit cell consisting of two columns of atoms of width $b=a$ [green rectangle in Fig~\ref{Fig1}(b) ], this results in the band structure of Fig.~\ref{Fig1}(a). Now, if a perturbation with \textit{hexagonal} periodicity like the Kekul\'{e} distortion is considered, a tripled sized unit cell (of width $b=3a$) is required to describe the whole system [red rectangle in Fig\ref{Fig1}(d)]. Taking such unit cell results in a three-folding in $k$-space (respect with that of pristine ZGNR) of the energy subband structure \cite{fujimoto2013} as shown in Fig.~\ref{Fig1}(c). The colors have been added to  better illustrate such  folding.  
 
 The characterization of the electronic conductance is performed within the Landauer regime of quantum transport. All the numerical calculations of the  bandstructure  and electronic conductance were done  using the Kwant platform \cite{kwant-paper} and subsequently compared and tested with the results of our own implemented codes following the standard recursive Green function method\cite{mucciolo-disorder}.

\section{Results} \label{sec:Results}

\begin{figure}[!htbb]
\begin{center}
\includegraphics[scale=0.152]{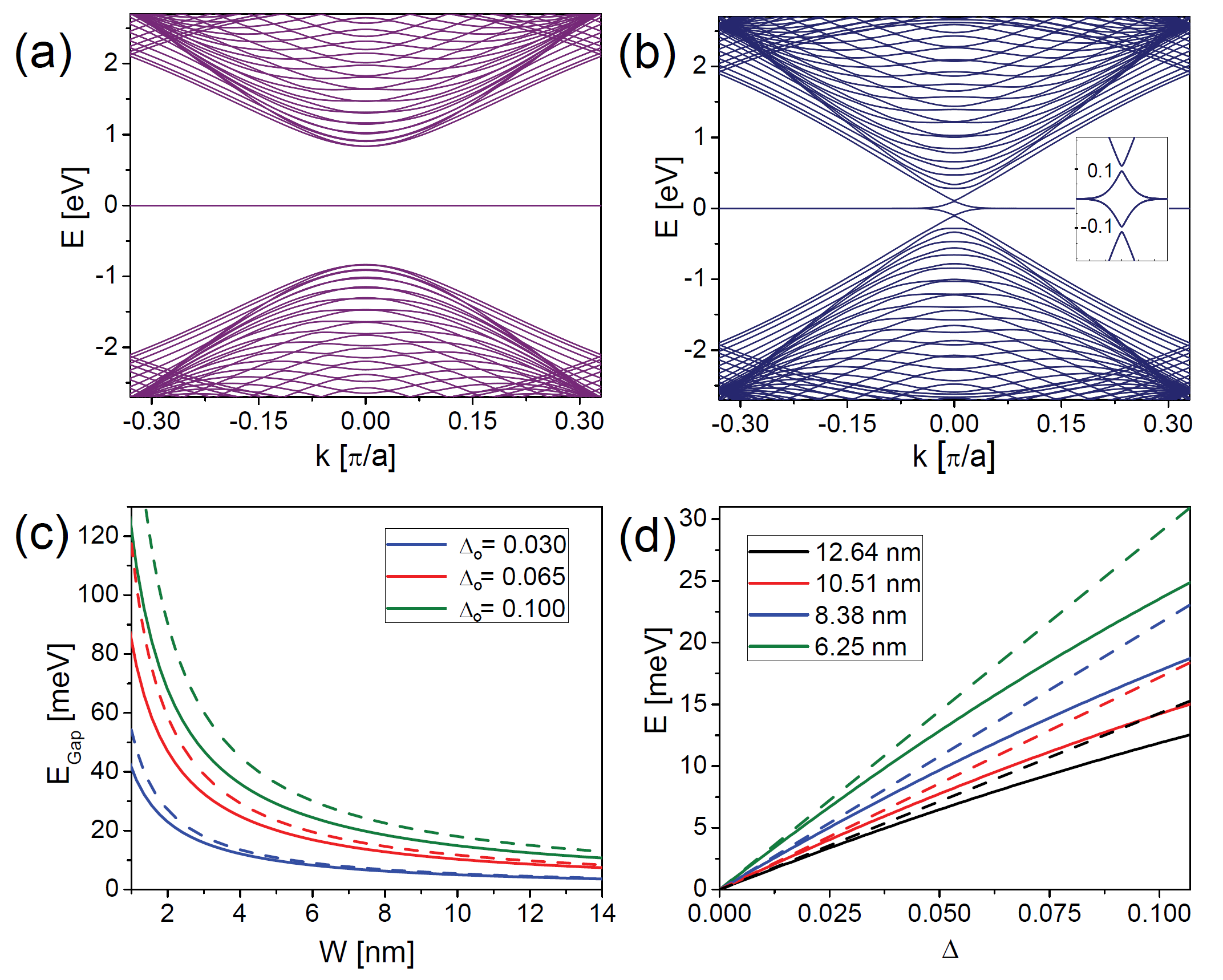}
\end{center}
\caption {Band structure for zigzag nanoribbons of (a) Kek-O distorted graphene, (b) Kek-Y distorted graphene where we considered widths of around 8.38 nm and $\Delta_0=0.1$. The size of the gap opened for ZGNR with Kek-Y distortion as function of (c) the nanoribbon's width, $W$  and (d) the Kekul\'e coupling amplitude $\Delta_0$  ; solid lines indicate the numerical results obtained through the tight-binding model, while dashed lines correspond to our analytic results in the low-energy approximation.
\label{Fig2}}
\end{figure}

Depending upon the type of Kekul\'{e} distortion  the effect on the band structure can be dramatically different. For instance, it is known that in an infinite  sheet of graphene, the  Kek-O distortion generates an energy gap that scales as   six times the Kekul\'e parameter $\Delta$\citep{chamon2000solitons,gamayun2018valley}. For the case of zigzag edge graphene nanoribbons, the Kek-O distortion also opens a gap\cite{armchair-kek} and, as shown in Fig.~\ref{Fig2}(a), a flat band persist at zero energy\cite{kariyado2017topological}. As a consequence these states do not contribute to  the electronic conductance. In contrast, a Kek-Y distortion in an infinite sheet of graphene does not lead to the aperture of a gap, however small energy gaps emerges at zero momentum in zigzag graphene nanoribbons, see inset in Fig.~\ref{Fig2}(b). The appearance of these gaps is somewhat surprising considering the fact that that, neither the Kek-Y distortion for infinite graphene flake, nor the zigzag cutting edge open an energy gap by themselves, however the combined scenario, Kek-Y in a ZGNR does yields energy gaps as shown here in inset of Fig.~\ref{Fig2}(b). Such gaps can be further characterized using the continuous approximation, as discussed below.

\subsection{Kek-Y Zigzag Graphene Nanoribbons}
Here we analyze the nature and magnitude of the gap for  Kek-Y distorted graphene nanoribons in  the low energy approximation. In this regime its Hamiltonian has the form\cite{gamayun2018valley},
\begin{equation}
H=
v_0 |p|
\begin{pmatrix} \label{Ec:Hyzz}
    0       & -e^{-i\theta} & -\Delta_0 e^{-i\theta}  & 0 \\
    -e^{i\theta}     & 0 & 0 & \Delta_0 e^{-i\theta} \\
    -\Delta_0 e^{i\theta}      & 0 & 0 & e^{-i\theta} \\
    0 & \Delta_0 e^{i\theta} & e^{i\theta} & 0 \\
\end{pmatrix}
,
\end{equation}
where $ |p|e^{\pm i \theta}= p_x \pm i p_y$ characterizes the electronic momentum, $v_0$ is the Fermi velocity of the carriers in freestanding graphene, and $\Delta=\Delta_0$ describes the strength of the Kekul\'e distortion. Setting the $x$-axis parallel to the nanoribbon we can write the wave function solution of Eq. (\ref{Ec:Hyzz}) as,
\begin{equation}\label{Eq:ansatz}
\Psi = \begin{pmatrix}
\phi_B'(y) \\
\phi_A'(y) \\
\phi_A(y) \\
\phi_B(y) \\
\end{pmatrix}
e^{ik_xx},
\end{equation}
in which the subindex $A,B$ denotes the sublattice, whiles the prime and unprime superindexes refers to the $K',K$ valleys, respectively. After substituting Eq. (\ref{Eq:ansatz}) into the  Dirac-like equation, $H\Psi=E\Psi$, it can be shown that the resulting differential equations should be satisfied by transverse wave functions of the explicit form $\phi_S(y)=c_{S}e^{\lambda_E(k_x) y}$, where $S=\{A,B\}$, and $c_{S}$ is a real constant; and similarly for the primed solutions. This leads to a quartic algebraic equation for $\lambda_E(k_x)$, where its four roots are given by $\pm \lambda_{+} , \pm\lambda_{-}$ with  

\begin{equation}
\lambda_{\pm}=\pm\sqrt{k_x^2-\frac{\epsilon^2}{\Delta_\pm^2}},
\label{Eq:lambda}
\end{equation}
where we have set $\epsilon=E/\hbar v_0 $ and $\Delta_{\pm}=1 \pm \Delta_0$. Therefore, the most general functional form for the transverse wave functions must be governed by 
\begin{subequations}
\begin{equation}
\phi_A(y)=Ae^{\lambda_- y}+Be^{- \lambda_-y }+Ce^{\lambda_+ y}+De^{- \lambda_+y },
\end{equation}

\begin{equation}
\phi_A'(y)=-Ae^{\lambda_- y}-Be^{-\lambda_- y}+Ce^{\lambda_+ y}+De^{-\lambda_+ y},
\end{equation}

\begin{equation}
\phi_B(y)=\frac{1}{\epsilon}[\gamma^{+}_{-}Ae^{\lambda_- y}+\gamma^{-}_{-}Be^{-\lambda_- y}+\gamma^{+}_{+}Ce^{\lambda_+ y}+\gamma^{-}_{+}De^{-\lambda_+ y}],
\end{equation}

\begin{equation}
\phi_B'(y)=\frac{1}{\epsilon}[\gamma^{-}_{-}Ae^{\lambda_- y}+\gamma^{+}_{-}Be^{-\lambda_- y}-\gamma^{-}_{+}Ce^{\lambda_+ y}-\gamma^{+}_{+}De^{-\lambda_+ y}],
\end{equation}
 
\end{subequations}

\noindent where $\gamma^{\alpha}_{\eta}=\Delta_\eta(k_x+\alpha\lambda_\eta)$, being $\alpha,\eta\in(+,-)$. Applying boundary conditions at the edges of the nanoribbon, $\phi_A(W)=\phi_A'(W)=\phi_B(0)=\phi_B'(0)=0$, we get the set of equations, 
\begin{equation}
\begin{pmatrix}
e^{\lambda_- W} & e^{-\lambda_- W} & e^{\lambda_+ W} & e^{-\lambda_+ W}\\
- e^{\lambda_- W} & -e^{-\lambda_- W} & e^{\lambda_+ W} & e^{-\lambda_+ W}\\
\gamma^{+}_{-} & \gamma^{}_{-} & \gamma^{+}_{+} & \gamma^{-}_{+}\\
\gamma^{-}_{-} &\gamma^{+}_{-} & -\gamma^{-}_{+} & -\gamma^{+}_{+}
\end{pmatrix}
\begin{pmatrix}
A \\
B \\
C \\
D \\
\end{pmatrix}
= \begin{pmatrix}
0 \\
0 \\
0 \\
0 \\
\end{pmatrix},
\end{equation}
from which the non-trivial solution should satisfy, 
\begin{equation}
\text{tanh}(\lambda_+ W)\text{tanh}(\lambda_- W)=\frac{\lambda_+\lambda_-}{k_x^2},
\label{Eq:Condition}
\end{equation}
being an implicit equation for the energy $E$ as a function of the momentum $k_x$. Then the gap of interest can obtained from Eq.~(\ref{Eq:Condition}). By setting  $k_x = 0$ in Eq.~(\ref{Eq:Condition}) we get the condition,
\begin{equation}
\text{cosh}(\lambda_+ W)\text{cosh}(\lambda_- W)=0,\\
\end{equation}
with
\begin{equation} 
\lambda_+ = i\frac{m\pi}{2W}, \quad \lambda_- = i\frac{n\pi}{2W},
\label{Eq:lambdamn}
\end{equation}
where $m$ and $n$ are integers. Since the size of the gap is given by the states around $E=0$ only the solutions for $m=n=1$ are required. Therefore, from Eq. (\ref{Eq:lambdamn}) and Eq. (\ref{Eq:lambda}) we get

\begin{equation}
E_\pm = \frac{\hbar v_0 \pi}{2W} \Delta_\pm,
\end{equation}
for the lowest  positive energies. The size of the energy gap is then given by,
\begin{equation}
E_ {Gap} = E_+ - E_- = \frac{\hbar v_0 \pi}{W} \Delta_0,
\label{Egap}
\end{equation}
which is also valid for the gap at negative energies [inset in Fig.~\ref{Fig2}(b)]. Plots of gap magnitude $E_{Gap}$ as a function of the nanoribbon width $W$ and the Kek-parameter $\Delta$ are shown in Fig.~\ref{Fig2}(c) and Fig.~\ref{Fig2}(d), respectively. The continuous lines are the plots of the analytic results from Eq.~(\ref{Egap}) and the dashed lines the corresponding numerical solutions. We notice that, as expected from a continuous model, the coincidence between both results is better for wider nanoribbons and for small values of the Kek-parameter $\Delta_0$. Similarly, the energy at which the gap is centered is given by
\begin{equation}
E^0_{gap}=\frac{1}{2}(E_+ + E_-)= \frac{\hbar v_0 \pi}{2W},
\end{equation}
which is independent of the Kekul\'{e} parameter.

\subsection{Kek-O Zigzag graphene nanoribbons}
A similar approach can be followed to find an expression for the gap in Kek-O textured GNR. The corresponding low energy Hamiltonian for the Kek-O distortion in graphene is given by, 

\begin{equation}
H=
\begin{pmatrix}
    0       & -v_0 |p| e^{-i \theta} & -3\Delta_0 t & 0 \\
    -v_0 |p| e^{i \theta}  & 0 & 0 & -3 \Delta_0 t \\
    -3 \Delta_0 t  & 0 & 0 & v_0 |p| e^{-i \theta} \\
    0 & -3 \Delta_0 t & v_0 |p| e^{i \theta} & 0 \\
\end{pmatrix}
,
\end{equation}
with a wavefunction of the generic form as given by Eq. (\ref{Eq:ansatz}) and Eqs. (8a-8d), but instead of Eq.(7) we find for this case,
\begin{equation}
\lambda=\sqrt[]{k_x^2+(3\Delta_0 t)^2-\epsilon^2},
\end{equation}
($\epsilon=E/\hbar v_0 $) and obtain a set of linear equations with coefficients, 

\begin{equation} \label{Eq:MkekO}
\begin{pmatrix}
e^{\lambda L} & e^{-\lambda_ L} & 0 & 0\\
0 & 0 & e^{\lambda L} & e^{-\lambda L}\\
(k_x+\lambda) &(k_x-\lambda) & -3\Delta_0 t & -3 \Delta_0 t\\
3\Delta_0 t & 3\Delta_0 t & (k_x-\lambda) & (k_x+\lambda)
\end{pmatrix}.
\end{equation}
requiring that the non-trivial solution satisfy,
\begin{equation}
\text{cosh}(2 \lambda W) = \frac{k_x^2+(3 \Delta_0 t)^2+\lambda^2}{k_x^2+(3 \Delta_0 t)^2-\lambda^2},
\label{Eq:KekOBands}
\end{equation}

\noindent which can be solved numerically to obtain the value of the gap as function of the nanoribbons width and the Kek parameter $\Delta_0$. In analogy with the former case (Kek-Y) and by setting $k_x=0$, we can explicitly show the value of the band gap, from Eq. (\ref{Eq:KekOBands}) we can write,
\begin{equation}
    \epsilon^2\left[1+\text{cosh}\left(2W \sqrt{(3 \Delta_0t)^2-\epsilon^2}\right)\right]=2(3 \Delta_0 t)^2,
\end{equation}
where the first solutions are given by $\epsilon=\pm 3 \Delta_0t$. Therefore, the size of the gap is given by $6 \Delta_0 t$, which coincides with the gap for an infinite Kek-O distorted graphene sheet\cite{Chamon2007,gamayun2018valley}. This last result applies only to zigzag GNR, for the Kek-O armchair GNRs the gap turn to be  considerable smaller\cite{armchair-kek,lin2017competing}. Notice that there is an entirely flat band at zero energy, this band will not contribute to transport since it has vanishing group velocity\cite{Wakabayashi2001}.             

\subsection{Quantum conductance calculation}

The electronic conductance is obtained using the wave function formulation of the quantum scattering problem\cite{kwant-paper} within the Landauer framework of ballistic transport\cite{Landauer} at zero temperature. It involves the calculation of the electronic transmission probabilities $T(E)$, through the whole nanoribbon attached to ideal (pristine) graphene contacts; the conductance is then obtained through the expression $G(E)=\frac{e^2}{h}T(E)$, where $T(E)$ comprises the sum of all transmitted modes allowed. Unless stated otherwise, we use  uniform regions of Kek-Y distorted graphene nanoribbons of width  $W=28.87\,a\simeq7.1 \text{ nm}$ and length $L=100.5\,a\simeq24.7 \text{ nm}$, in a sequentially ordered pattern with alternating orientations  ($\nu=\pm1$).

\begin{figure}[!htbp]
\begin{center}
\includegraphics[scale=0.30]{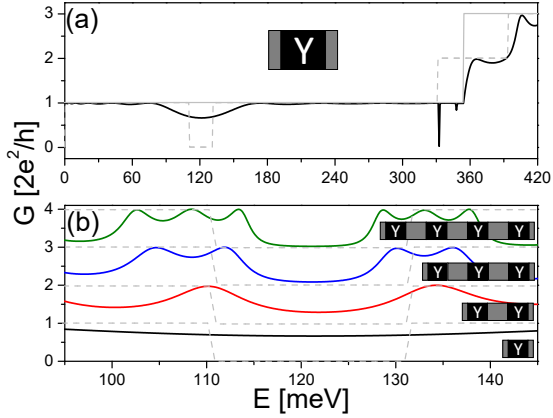}
\end{center}
\caption {(Color online) (a) Conductance of system compound of a Kek-Y distorted graphene slab with pristine graphene leads (black line), conductance of pristine graphene (solid gray line) and Kek-Y distorted graphene (dashed gray line) are also shown. (b) Conductance of systems of alternating slabs of Kek-Y distorted graphene and pristine graphene for one slab (black line), three slabs (red line), five slabs (blue line) and seven slabs (green line). Notice that the real physical spin does not any role here, therefore the conductance is spin-degenerate, which explains the factor of two in the units of conductance. 
} 
\label{Fig3}
\end{figure}

The black solid line in Fig.~\ref{Fig3}(a) shows the electronic conductance as a function of the Fermi energy through a single slab of Kek-Y ($\nu=+1$) distorted Zigzag graphene nanoribbon. The gray dashed line is the conductance for an infinite Kek-Y distorted GNR, and for comparison, we have plotted also (solid gray line) the  conductance in the absence of any Kek-distortion ({\it i.e.} for pristine GNR). We notice that for the presence of the Kek-Y distortion everywhere (including the leads), the zero step in the conductance presents a small gap close to $20\, \text{meV} $   at the incident energy centered at about $E^0_{gap}=125\, \text{meV}$   [gray dashed line in Fig.~\ref{Fig3}(a)]. Interestingly, a depletion of the conductance, centered at this energy, occurs when the distortion is present in a finite region of length $L$ (black solid line). Such dip in the conductance becomes more pronounced as the length of the distorted region increases (not shown). The black curve in Fig.~\ref{Fig3}(a) also shows Fabry–Pérot oscillations, as the Fermi energy is increased, associated with the finite size of the Kek-Y distorted region\cite{Schomerus}. Another interesting feature is the splitting in two of the first step in the conductance ($E_{1st}\approx340 \text{ meV}$). This is associated with the breaking of the valley degeneracy created by the Kek-Y distortion which couples both valleys\cite{gamayun2018valley} at a given momentum $k$, which for an infinite graphene sheet, can be evaluated by degenerated perturbation theory, and results in a displacement in energy of the cones by $\pm\Delta_0 E$. Therefore the expected energy difference of the new step is $2\Delta_0 E_{1st}\approx68\text{ meV}$. This splitting is a signature of a Kekul\'{e}-Y distortion  in GNRs and its measurement provides an indirect and independent way to estimate the strength of the Kekul\'{e} parameter. The conductance dip observed around $E\approx330 \text{ meV}$ is most likely due to  a resonant backscattering by a quasi-bound state in the distorted region caused by the confinement with  the leads. We notice that the appearance of such resonances occurs as long there are two available modes in the central kek-Y region while there is only one mode available in the leads\cite{GaussianPotential}. 

Now, by adding subsequent Kek-Y distorted regions, separated by pristine graphene slabs of the same length $L$, generates symmetric resonant peaks in the conductance, as shown in Fig.~\ref{Fig3}(b). The red line shows the conductance for two Kek-distorted regions (one pristine graphene slab between them) and it presents two resonant peaks approximately at $E^0_{\pm}=E^0_{gap}(1\pm\Delta_0)$, respectively. Clearly, the number of resonance peaks emerging around $E^0_{+}$ and $E^0_{-}$ correspond to the number of clean (no Kek-distorted) graphene slabs in the Kek-Y superlattice [See blue and green lines in Fig.~\ref{Fig3}(b)]. Equivalently, the number of resonance peaks goes as the number of Kek-Y slabs minus one.
Such resonant effect pretty much resembles the resonant tunneling transmission probability of semiconducting 
multibarrier  structures\cite{ResonancesMultiRomo,Pereira1998}, in which the Kek-Y regions plays the role of the physical potential barriers. However, in contrast to the latter, the resonances and its multiplicity appears in Kek-Y/Graphene/Kek-Y superlattices in the absence of band-offsets and/or local gating. 

\begin{figure}[!htbp]
\begin{center}
\includegraphics[scale=0.32]{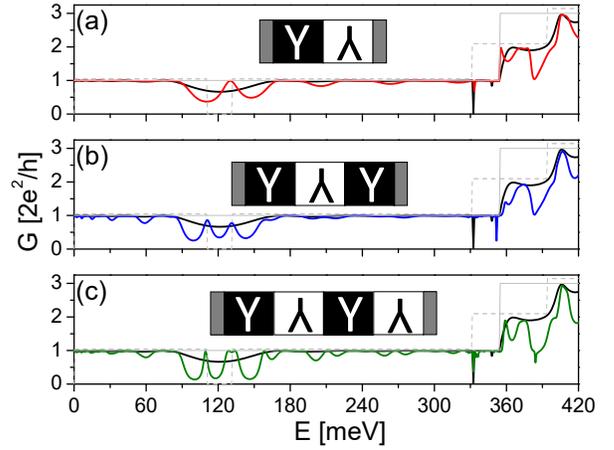}
\end{center}
\caption {(Color online) Conductance of systems compound of slabs with alternating orientation of Kek-Y distorted graphene and pristine graphene leads for (a) two slabs (red line), (b) three slabs (blue line), and (c) four slabs (green line). The conductance of pristine graphene (solid gray line), Kek-Y distorted graphene (dashed gray line) and one Kek-Y slab (black line) are also shown.}
\label{Fig4}
\end{figure}

\begin{figure}[!htbp]
\begin{center}
\includegraphics[scale=0.33]{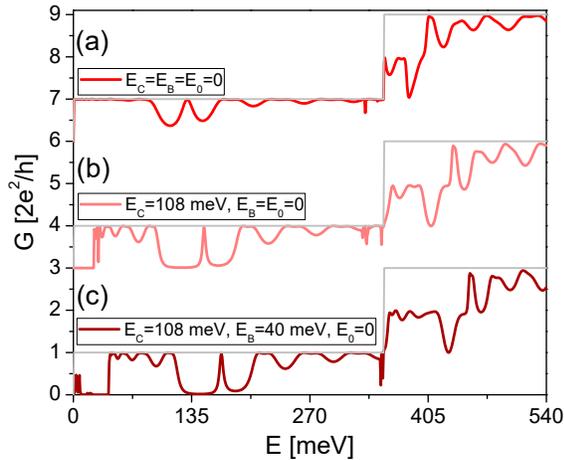}
\end{center}
\caption {(Color online) Conductance through two slabs of Kek-Y distorted graphene with opposites orientations considering: (a) without on-site energy modifications, (b) with on-site energy modifications $E_c=108 \text{ meV}$ for the sites surrounded by three the modified bonds and (c) on-site energy difference of $E_c=108 \text{ meV}$ and $E_b=40 \text{ meV}$ for the sites surrounded by three and one modified bonds, respectively.
\label{Fig5}}
\end{figure}

Next, we explore the conductance response for the case of Kek-Y distorted regions but with the opposite orientations $\nu=\pm1$. We observe that the single Kek-Y/Kek-\rotatebox[origin=c]{180}{Y} interface develop a resonance in the otherwise gap region (Fig.~\ref{Fig4}(a)).
As the number of alternating Kek-regions increases, the number resonances in the otherwise gap region also increases [see Fig.~\ref{Fig4}(a)-(c)].
It is known that a Kek-Y distortion in graphene will also generate an asymmetric on-site energy pattern due to the modifications of the relative distances between sites and the corresponding modification of the crystal field\cite{bardeen1950deformation,suzuura2002phonons,gutierrez2016imaging}. This behavior has been studied in a recent paper by J. J. Wang et al.\cite{Juan-prb}. In order to further study the persistence of the conductance resonances in the presence of this on-site energy asymmetry we performed the numerical calculations shown in Fig.~\ref{Fig5}. We focus our study to the conductance through two Kek-Y distorted graphene slabs with opposites orientations as in Fig.~\ref{Fig4}(a). For reference we plot (Fig.~\ref{Fig5}(a)) the conductance without on-site energy modifications [as in Fig.~\ref{Fig4}(a)]. The case when taking different on-site energy  $E_c=108 \text{ meV}$ for the sites surrounded by three modified bonds (as in the work by Wang et al.\cite{Juan-prb} ) is depicted in Fig.~\ref{Fig5}(b).  The most asymmetric case is obtained by considering on-site energy difference of $E_c=108 \text{ meV}$ or $E_b=40 \text{ meV}$ for the sites surrounded by three or one modified bonds, respectively. To guide the eye we plotted in gray the conductance of an infinite pristine GNR. The conductance plots reveal that the resonant states persist in the presence of on-site energy modifications, the same happens form interfaces with more slabs (not shown here). Another consequence of the on-site energy modifications is shifting of the conductance response to higher energies, this evidently results in the increasing of the energy range for the appearance of gap and for the energy range of the first step in the conductance. 

\section{Conclusions} \label{sec:conclusions}
We have presented a first study of the quantum conductance of Kek-Y distorted zigzag graphene nanoribbons. Our results show that the breaking of the valley degeneracy due to Kek-Y distortions in infinite graphene is robust under confinement in the form of nanoribbons. The latter is reflected in the sub-band structure that results in a valley-resolved spectrum, and in the behavior of the electronic conductance. We show that the Kek-Y distortion generates a gap in the zeroth step for an infinite long Kek-Y distorted ZGNR. Using the continuum approximation we find a simple analytical expression for the size of this gap as function of the nanoribbons width and the Kek-Y strength-parameter. In addition, we presented transport studies of a ZGNR with regions of alternating Kek-Y distortions of the same spatial length.
The superlattice of Kek-Y interfaces show the appearance of resonant tunneling transport when two or more slabs of Kek-Y distorted regions with the same or different orientation are considered. Furthermore, we have found that including the asymmetrical on-site energy modifications to the Kek-Y distortions does not destroys these resonances. Considering that the Kek-Y distortion is an experimental reality, and as long uniform Kek-Y distortions could be manipulated in actual graphene nanoribbons, all these effects would be in the reach of experimental measurements, opening  the exciting possibility of designing resonant tunneling graphene-based devices.

\section*{Acknowledgments}
E.A. and R.C.-B. acknowledges useful discussions with Priscilla Iglesias, Mahmoud Asmar and Gerardo G. Naumis. The plots in Fig.\ref{Fig3}-\ref{Fig5} were created using the software Kwant \citep{kwant-paper}.  This work was supported in part by project PAPIIT-UNAM-IN111317. R.C.-B. thank 20va Convocatoria Interna (UABC). P.A.P. thank to Mexico’s National Council of Science and Technology (CONACYT) for their financial support.

\bibliographystyle{unsrt}
\bibliography{Refs.bib}

\end{document}